\documentclass[prb,preprint,showpacs,showkeys,superscriptaddress]{revtex4}

\usepackage{hyperref}
\usepackage[english]{babel}
\usepackage[T1]{fontenc}
\usepackage[latin1]{inputenc}
\usepackage{amsmath}
\usepackage{mathenv}
\usepackage{bm}
\usepackage{babel}
\usepackage[v2]{xy}
\usepackage{epsfig}
\usepackage{amsfonts}
\usepackage{amssymb}
\usepackage{color}

\begin{document}

\title{Parametric oscillator based on non-linear vortex dynamics in low resistance magnetic tunnel junctions.}
\date{\today}
\author{S. Martin}
\author{N. de Mestier}
\affiliation{SPINTEC, UMR-8191,CEA-INAC/CNRS/UJF-Grenoble1/Grenoble-INP, 17 rue des martyrs, 38054 Grenoble Cedex 9, France}
\author{C. Thirion}
\author{C. Hoarau}
\affiliation{Institut Néel, CNRS et Université Joseph Fourier, BP 166, F-38042 Grenoble Cedex 9, France}
\author{Y. Conraux}
\affiliation{Crocus-Technology, 5 place Robert Schuman, 38025 Grenoble cedex, France}
\author{C. Baraduc}
\author{B. Diény}
\affiliation{SPINTEC, UMR-8191,CEA-INAC/CNRS/UJF-Grenoble1/Grenoble-INP, 17 rue des martyrs, 38054 Grenoble Cedex 9, France}

\begin{abstract}

Radiofrequency vortex spin-transfer oscillators based on magnetic tunnel junctions with very low resistance area product were investigated. A high power of excitations has been obtained characterized by a power spectral density containing a very sharp peak at the fundamental frequency and a series of harmonics. The observed behaviour is ascribed to the combined effect of spin transfer torque and Oersted-Ampère field generated by the large applied dc-current. We furthermore show that the synchronization of a vortex oscillation by applying a ac bias current is mostly efficient when the external frequency is twice the oscillator fundamental frequency. This result is interpreted in terms of a parametric oscillator.

\end{abstract}
\maketitle
\clearpage

\section{introduction}
The interaction between a spin-polarised current and a local magnetization, via spin transfer torque, can induce steady-state magnetization precessions. This effect is intensively studied both from a fundamental point of view as well as for its potential application as radiofrequency oscillators. Two classes of these so-called spin torque oscillators (STO) are being studied. In a first class, the magnetization of the excited layer is close to single domain and precesses around its equilibrium position. In the second class, magnetization forms a magnetic vortex \cite{Pribiag2007, Mistral2008, Finocchio2008} that is driven into a periodic circular motion by spin transfer torque. In both cases, improving the output power and the quality factor of these STO is a challenging requirement for practical applications of these devices. In this context, we studied vortex oscillators based on magnetic tunnel junctions that have a high output power up to $20~nW$. 
This output power might be further enhanced by synchronizing several oscillators. This concept has been tested by using either magnetic coupling \cite{Kaka2005} or direct serial or parallel electrical connexion \cite{Georges2008synchronization}. It was also shown that a STO could be locked on an external source at a frequency close to its natural frequency $f_0$, either by feeding the oscillator with a ac bias current \cite{Rippard2005, Georges2008locking, Lehndorff2010} or by applying a RF magnetic field \cite{Bonin2009}. Theoretically, locking should also be possible at any frequency being a rational multiple of the natural frequency \cite{Pikovsky2005}. This was experimentally demonstrated on macrospin STO \cite{Urazhdin2010fractional}.
In this paper we show however that synchronization of our vortex oscillator is not obtained for many frequencies. In particular, the most efficient synchronization is observed when applying a ac bias current at a frequency equal to twice the fundamental frequency. This result is explained by considering the complete system formed by the vortex oscillator and the external microwave source as a parametric oscillator.\\

\section{Sample characterization}
Our samples are ultra-low RA magnetic tunnel junctions (MTJ) in which vortex oscillations can be observed when the junction is subjected to a large dc bias current and a low in-plane field. The large dc current produces both a large Oersted-Ampère field which, together with the in-plane demagnetizing field, is responsible for the vortex nucleation and a spin transfer torque that starts the vortex oscillation. A similar behaviour was observed in low RA MTJ subjected to out-of-plane polarized spin current \cite{Dussaux2010}. In contrast to these experiments, the vortex oscillations are observed in our case with in-plane polarized spin current.
Our MTJ are composed of the following layers: $IrMn_7 /CoFe_2 /Ru_{0.7} /CoFe_{2.5} /AlOx/CoFe_3/NiFe_5$ where the subscript corresponds to the layer thickness in nm. After deposition, the stack was etched into a pillar with a $300~nm$ diameter circular cross section. The MR ratio is about 12\% and the resistance area product is equal to $0.3~\Omega.\mu m^2$. In such low RA junctions, there may be some spatial inhomogeneities in the current flow due to hot spots in the barrier as suggested in the work of Houssameddine et al. \cite{Houssameddine2008}. 
\begin{figure}[htb]
	\includegraphics[width=\columnwidth]{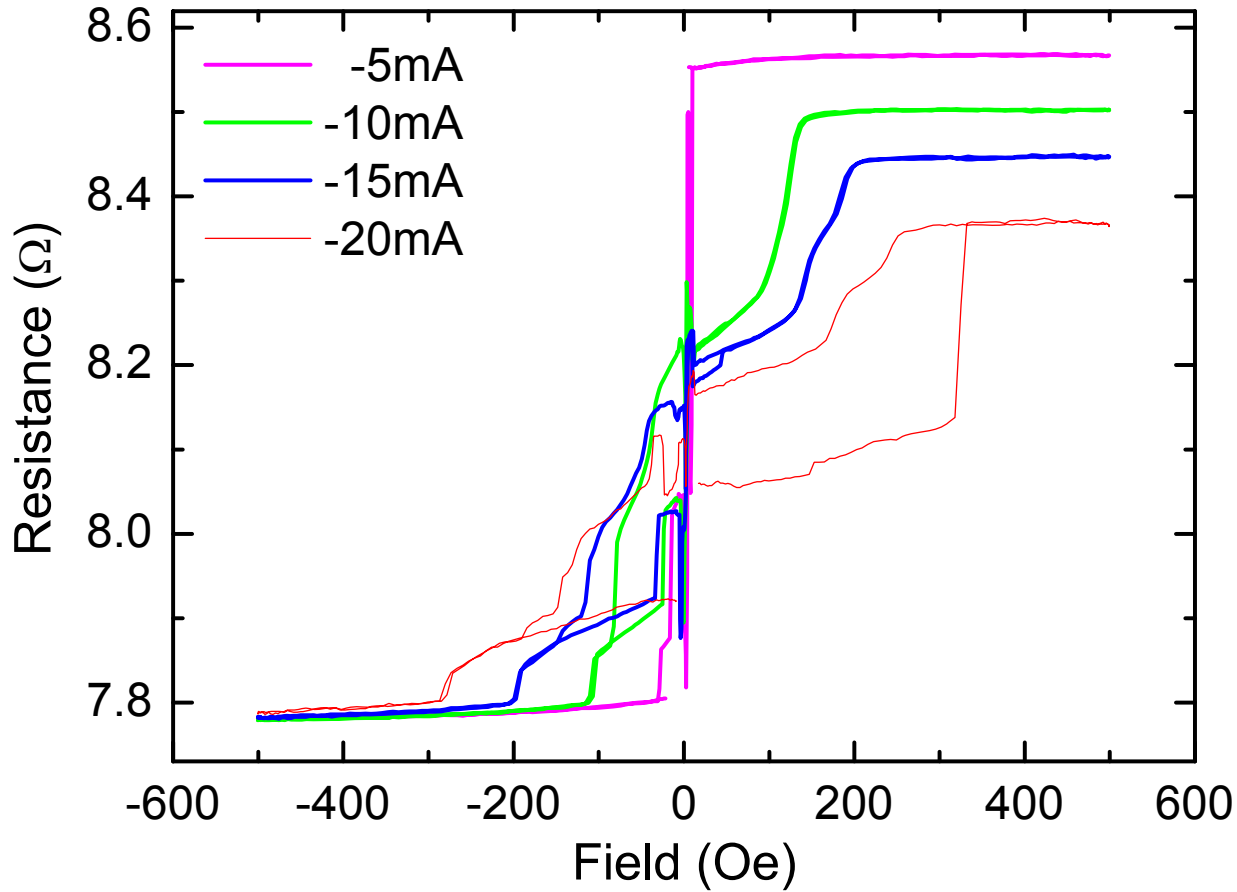} 
	\caption{(Color online) Magnetoresistance as a function of the applied field for different currents. The curves are vertically shifted in order to overlap in the parallel state. The field is applied in-plane along the easy axis. The sample is prepared in the parallel state ($-600~Oe$) before the field cycle. The cycle starts from the parallel state to the antiparallel state and back to the parallel state. The electrons flow from the reference to the free layer. Similar behaviour is obtained for positive current.}
	\label{fig:RH}
\end{figure}
Four point static resistance measurements were performed on the samples as a function of the in-plane magnetic field applied along the easy-axis (R-H curves). At low current, the usual sharp transition is observed between the low resistance state (parallel state) and the high resistance state (antiparallel state). At higher current value ($\left|I\right|>5~mA$),  the R(H)  hysteresis loops become more slanted and steps and hysteresis appear (Fig.~\ref{fig:RH}). These features become broader when the bias current is increased. This observation is interpreted as the formation of a vortex due to the Oersted-Ampère field generated by the current. This assumption is supported by the fact that our R-H curves are similar to those observed in thick circular samples where the stable micromagnetic state is a vortex \cite{Cowburn1999}. The vortex state is characterized by an average magnetization close to zero and thus by a device resistance equal to the mean value between the resistance values in parallel and antiparallel magnetic configuration. By increasing the applied field, we observed first a gentle slope of the curve that is ascribed to the vortex core moving toward the edge of the sample and then a sharp transition corresponding to the vortex expulsion at large field. The larger the applied current, the larger the Oersted-Ampère field and the more stable the vortex state. As a result, the low field slope of the R(H) curves decreases and the field value corresponding to vortex expulsion increases with bias current. Some hysteresis is observed on the curves at high current indicating that besides this overall picture, more complex detailed micromagnetic configurations may appear in these pillars. Nevertheless, the R-H curves although being dependent on the magnetic preparation of the sample before measurement are reproducible provided they are measured in the same conditions. The variation of $R_{max}$ as a function of current is due to the large bias dependence of the resistance in antiparallel configuration.\\

\section{spectral measurements}
\begin{figure}[htb]
	\includegraphics[width=0.95\columnwidth]{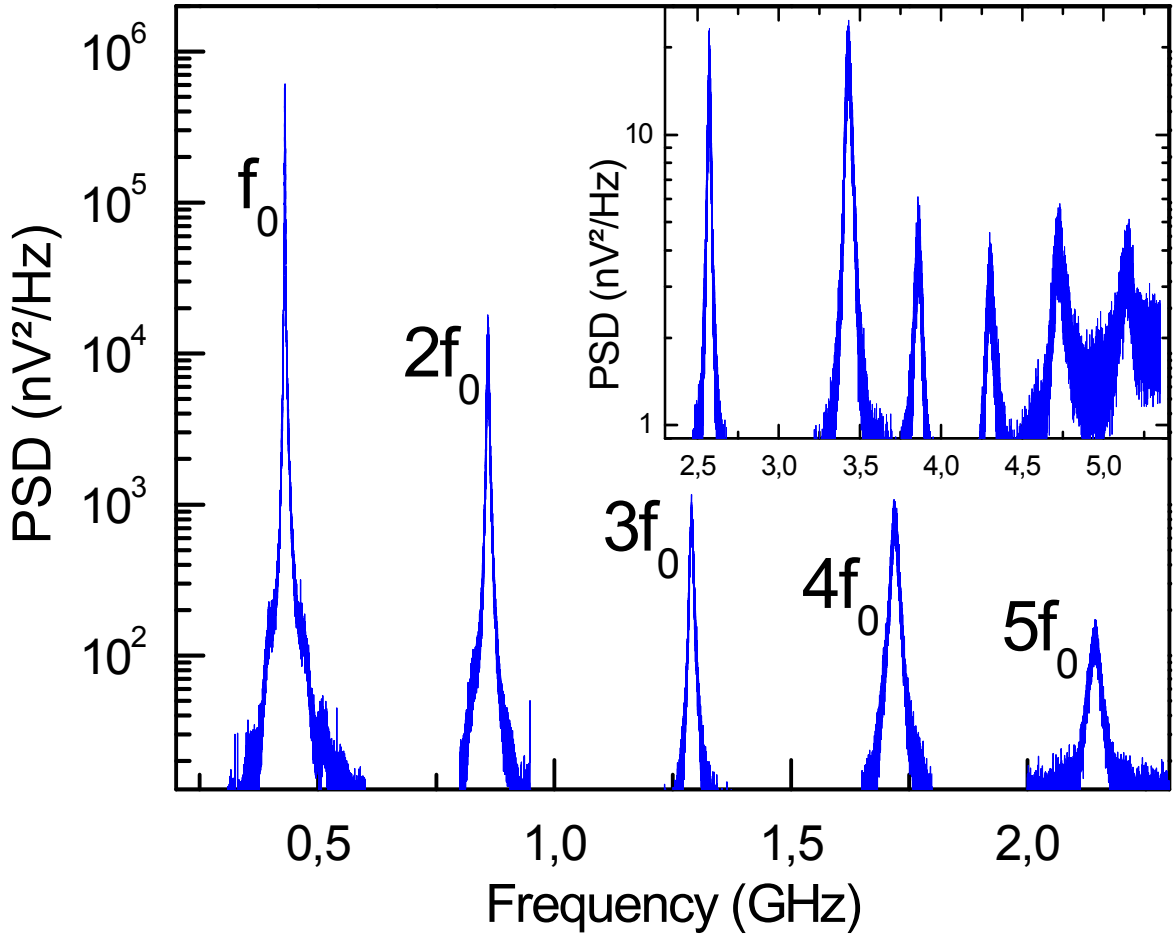} 
	\caption{(Color online) Power spectral density emitted by the junction in which the pre-existing static vortex is driven into motion. Measurements are performed with a bias current of $-20~mA$ and under $110~Oe$ in-plane applied field. Each peak was measured separately for a better resolution: harmonics 1 to 5. Inset: harmonics 6 and 8 to 12 (the 7th harmonic could not be detected).}
	\label{fig:spectrum}
\end{figure}

Spectral measurements were also performed on our samples. The MTJ is polarized by a dc current through a bias tee and the output power spectral density is measured with a spectrum analyzer. For bias currents larger than $17~mA$ (i.e.\ $2.4\;10^7 A/cm^2$) in absolute value, a radiofrequency spectrum composed of one large and narrow peak at the fundamental frequency $f_0$ and of about 10 harmonics of much smaller amplitude was observed (Fig.~\ref{fig:spectrum}). The amplitude of the largest peak is up to $1~\mu V^2/Hz$ in the best samples and its linewidth is about $1~MHz$ which corresponds to an integrated power of $20 nW$. The fundamental frequency varies from sample to sample between $350~MHz$ and $500~MHz$ and the relative amplitudes of the various peaks differ from sample to sample. The harmonic $2f_0$ is always much larger than the other harmonics and its voltage amplitude is about 10 times smaller than the fundamental peak. Its linewidth is twice larger than the linewidth of the fundamental peak. In fact, we more generally observed that the linewidth of the $N$-th harmonic $Nf_0$ is $2^N$ times larger than the fundamental peak linewidth.
When varying the static field or the dc-current, the peak frequency hardly changes showing only a slight increase of frequency with increasing the bias current ($~10MHz/mA$). This poor tunability, compared to macrospin STO, was already observed in nanocontact-based vortex oscillators \cite{Finocchio2008,VanKampen2009,Pufall2007}. In addition a strong hysteretic behaviour is also observed while sweeping the dc-current. When decreasing the bias current (in absolute value), the dynamical vortex state survives even at current smaller that the critical current $I_{on}$ corresponding to the onset of the dynamical state. The observation of two threshold currents $I_{on}$ and $I_{off}$  that correspond respectively to switching on and off the dynamical state was already reported in the case of nanocontacts \cite{Pufall2007}. In our samples, $I_{on}$ and $I_{off}$ are respectively about $17~mA$ and $6~mA$ (i.e.\ $2.4\; 10^7$ and $8.4\; 10^6 A/cm^2$).\\
The onset of a vortex dynamical state due to spin transfer torque is well understood in the case of out-of-plane spin polarized current \cite{Mistral2008}. This case has been extensively studied in the literature where most experiments are performed with out-of-plane applied field \cite{Dussaux2010,Mistral2008,Pufall2007}. In our experiment however, magnetic field is applied in-plane. In this case, it was pointed out that no vortex motion is expected for in-plane spin-polarized current, except in the case of a non uniform magnetic distribution in the polarizer \cite{Khvalkovskiy2010}. In our case, the Oersted-Ampère field is quite large ($250~Oe$ for $I_{dc}=20~mA$) compared to the RKKY coupling field estimated to $500~Oe$ assuming a reasonable value of the coupling energy ($J_{RKKY}=1~erg/cm^2$). Therefore the magnetization distribution of the reference layer is probably non uniform, being either a C-state or a strongly off-centered vortex. This magnetic distorsion in the pinned layer may explain the hysteresis observed at $20 mA$ in Fig.~\ref{fig:RH}. This distribution is responsible for an inhomogeneously polarized spin current that exerts a torque on the vortex core in the free layer leading to a gyrotropic motion of the vortex. When the current is reversed, the C-state will be reversed despite the presence of the exchange bias field and the current sign will change, so the spin torque term remains rougthly the same. In fact, this is experimentally observed: the vortex dynamical state exists for both positive and negative current. Finally, let us gain some insight into the vortex trajectory. Considering the numerous harmonics observed in the measured spectrum, the vortex orbit may be a more complex trajectory than a mere circle, such as an irregular ellipse along which the vortex core travels at non-constant velocity. From the total power of the first peak in the power spectrum, the resistance change during a period may be estimated and compared to the magnetoresistance measured for the corresponding dc-bias current; it is about 10\% of the maximum possible resistance change. So it is reasonable to assume that the area enclosed by the vortex trajectory in the free layer represents 10\% of the total surface of the sample, which means an average orbit radius of 1/3 of the sample radius.\\

\section{synchronization}
\begin{figure}[htb]
	\includegraphics[width=\columnwidth]{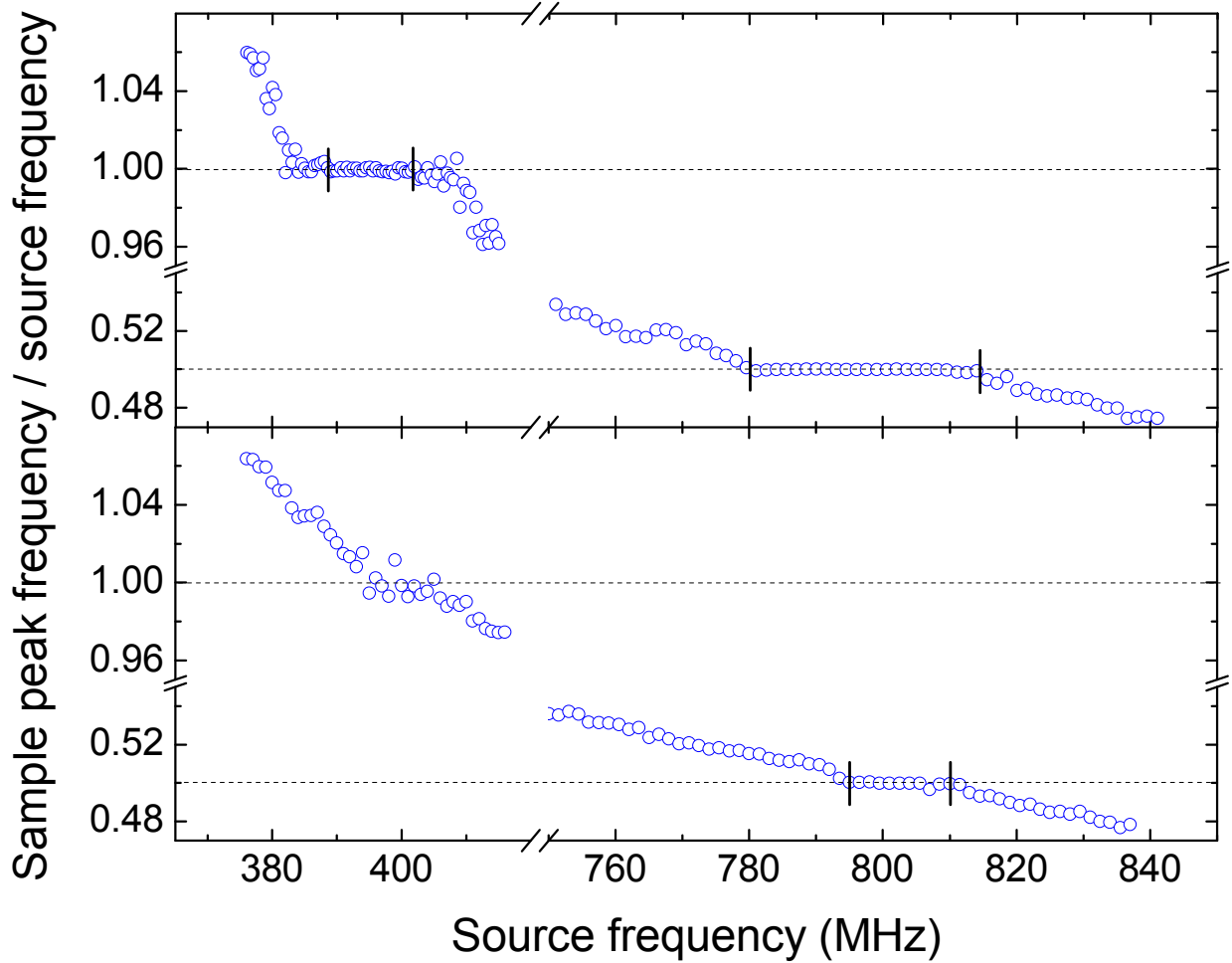} 
	\caption{(Color online) Normalized frequency of the fundamental peak versus frequency of the injected microwave; the power of the microwave is $-20~dBm$ in the upper panel ($I_{ac}~1 mA$) and $-28~dBm$ in the bottom panel ($I_{ac}~0.4 mA$); $I_{dc}=-20~mA$ and $H=250~Oe$.}
	\label{fig:locking}
\end{figure}

Synchronization of the oscillator with an external signal was then tested by applying an additional ac bias current to the magnetic tunnel junction. Since the sample is a one-port device, a power-splitter was used in order to connect the sample both to the spectrum analyzer and to the microwave source. The sample is therefore simultaneously subjected to a dc-current $I_{dc}$ and to a small ac-current $I_{ac}$. When sweeping the frequency $f_{ext}$ of the injected power, we observe that the oscillator fundamental frequency is tuned by the external signal. In Fig.~\ref{fig:locking} we show this phenomenon by sweeping the frequency $f_{ext}$ either around $f_0$, the oscillator fundamental frequency, or around $2f_0$. In the latter case, we observe that synchronization remains possible even for low injected power. Thus synchronization at twice the fundamental frequency appears to be most efficient. In that case, synchronization takes place on a range of a few tens of $MHz$ around the oscillator natural frequency which corresponds to a relative detuning range $\Delta f/f_0$ of about $4\%$. This modest detuning range may be directly related to the poor tunability of the vortex oscillator \cite{Georges2008locking}.

\begin{figure}[htb]
	\includegraphics[width=\columnwidth]{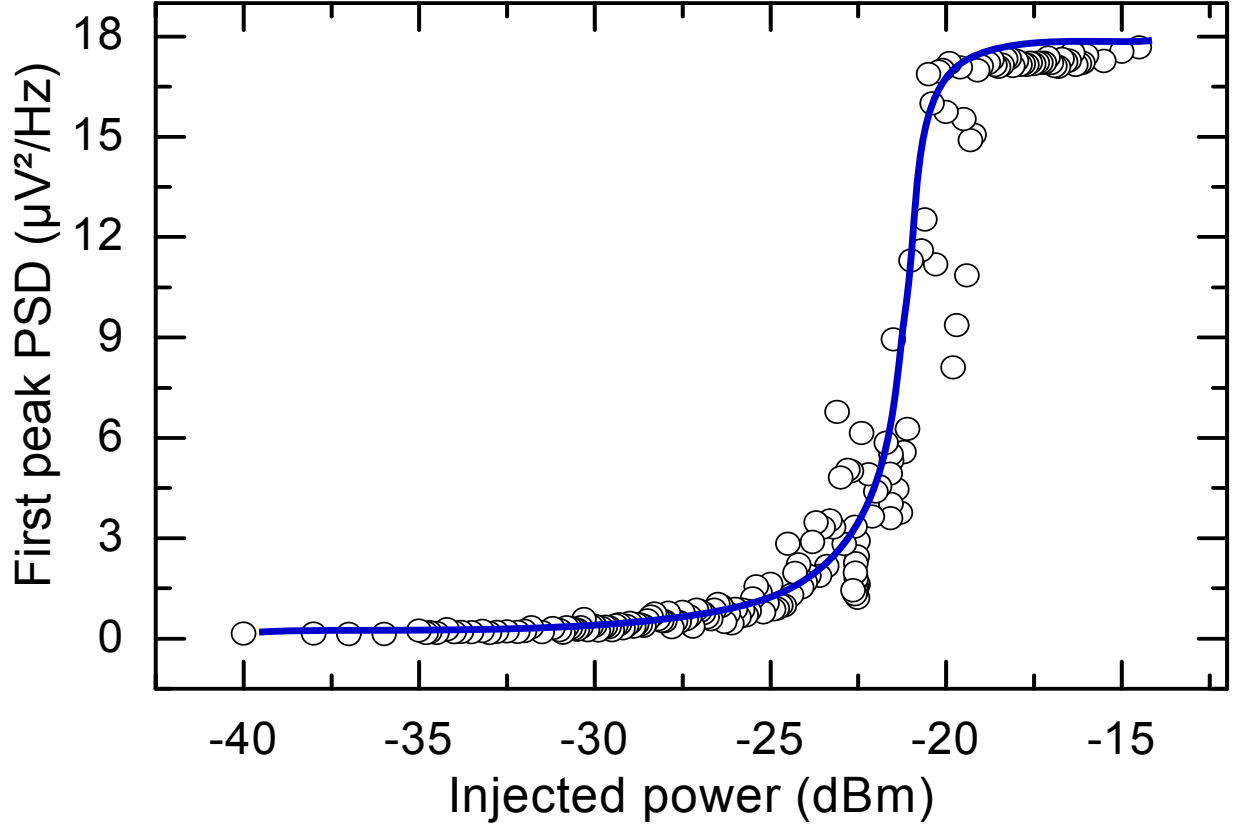} 
	\caption{(Color online) Amplitude of the fundamental peak versus power of the injected microwave at $2f_0$. $f_{ext}=700~MHz$; $I_{dc}=-17~mA$; in-plane field $H=15~Oe$. The blue line is only a guide for the eyes.}
	\label{fig:amplitude}
\end{figure}

Experimentally, synchronization at the oscillator fundamental frequency is observed on the spectrum analyzer as the superposition of the relatively broad peak of the oscillator and the very sharp peak of the microwave source. When synchronization is performed at twice the fundamental frequency, the oscillator peak is observed alone. Therefore it is possible to perform another experiment with keeping the external signal frequency fixed within the synchronization range and varying the input power. Fig.~\ref{fig:amplitude} shows the oscillator fundamental peak amplitude as a function of the input power. Above a critical input power value (about $- 25~dBm$ corresponding approximatively to $I_{ac}\approx0.5~mA<<I_{dc}$), the noise is drastically reduced around the fundamental peak and the amplitude of this peak increases by several order of magnitude. Simultaneously the peak linewidth becomes extremely narrow ($<1kHz$), whereas the integrated power is conserved.

\section{Model}
All our experimental observations may be explained by considering the equation of motion of the vortex subjected to the spin torque due to both the dc and ac bias currents. In order to take into account the time-dependent velocity of the vortex and consequent vortex distorsion, the vortex motion is described by the generalized Thiele equation \cite{Thiele1973} including the inertial term \cite{Wysin1996}:
\begin{equation}
M \ddot{\vec{X}}=\vec{G} \times \dot{\vec{X}} - \partial W/\partial \vec{X} - \eta \dot{\vec{X}} + \vec{F_{ST}}
\end{equation}
where $\vec{X}$ is the position of the vortex core, $M$ is the vortex mass, $\vec{G}=-G_0 \hat{z}$ is the girovector, $\eta$ is the damping constant and $-\partial W/\partial \vec{X}$ is the force due to the magnetostatic potential. Since the vortex is confined within the magnetic dot due to the strong Oersted-Ampère field, we choose a \textsl{ad-hoc} form of the potential. We assume that the potential has a parabolic shape in the center of the sample \cite{Guslienko2001} and confines the vortex within a limit radius $\xi$: $W(r)=\kappa \xi^2\left[2(1-r^2/\xi^2)\right]^{-1}$, so that the corresponding force may be written as $-\kappa f(r) \vec{X}$ where $f(r)=(1-r^2/\xi^2)^{-2}$. Finally, the effect of the spin transfer torque is described as a force $F_{ST}$ acting on the vortex and proportional to the total applied current $I(t)=I_{dc}+I_{ac} \cos(2 \omega t)$ (with $\omega_{ext}= 2\omega$), so that it may be written as $\vec{F_{ST}}=\lambda I(t) \hat{z}\times \vec{X}$, $\lambda$ being a proportionality factor \cite{Khvalkovskiy2010}. So the equation can be re-written in the following form:

\begin{eqnarray}
\label{mathieu}
M \ddot{\vec{X}}+ 
		\left[
		 \begin{matrix}
		  \eta & -G_0 \\
		  G_0 & \eta
		\end{matrix}
		\right]
 \dot{\vec{X}} + 
 \hspace{70pt} 
 \\	
		\left(
				\left[
		 		\begin{matrix}
		  		\kappa f(r) & \lambda I_{dc} \\ \notag
		  	 -\lambda I_{dc} & \kappa f(r)
				\end{matrix}
				\right]
				+ 
				\lambda I_{ac} \cos (2 \omega t)
				\left[
		 		\begin{matrix}
		  		0 & 1 \\
		  		-1 & 0
				\end{matrix}
				\right]
		\right)
  \vec{X} = \vec{0}
\end{eqnarray}

It is worth emphasizing that Eq.~(\ref{mathieu}) is a non-linear differential equation, by contrast to the work of Choi et al.\cite{Choi2009}, even when $I_{ac}$ is set to zero, because of the definition of the potential. Such type of equation is expected for a self-sustained oscillator. This equation with $I_{ac}=0$ may be solved assuming a circular trajectory of the vortex core with a radius $R$.
The critical current $I_c$ is defined as the minimum current that compensates the dissipation $\eta$. In that case, the oscillator pulsation $\omega_0$ is found to be a real number and the orbit radius is close to zero. When the bias current increases, the model predicts that the pulsation $\omega_0$ increases as well as the orbit radius. For sub-critical bias current, there is no steady-state motion of the vortex, since the pulsation $\omega_0$ is a complex number with an imaginary part describing the damping of the oscillation. All these predictions are qualitatively consistent with experimental observations. Quantitative comparison between model and experiment is more difficult due to the fact that some parameter values are not precisely known. Nevertheless, from the measured resonance frequency it was possible to estimate the vortex mass which is found of the order of $10^{-20}g$, consistently with previous theoretical calculation\cite{Guslienko2010}.

\begin{figure}[htb]
	\includegraphics[width=\columnwidth]{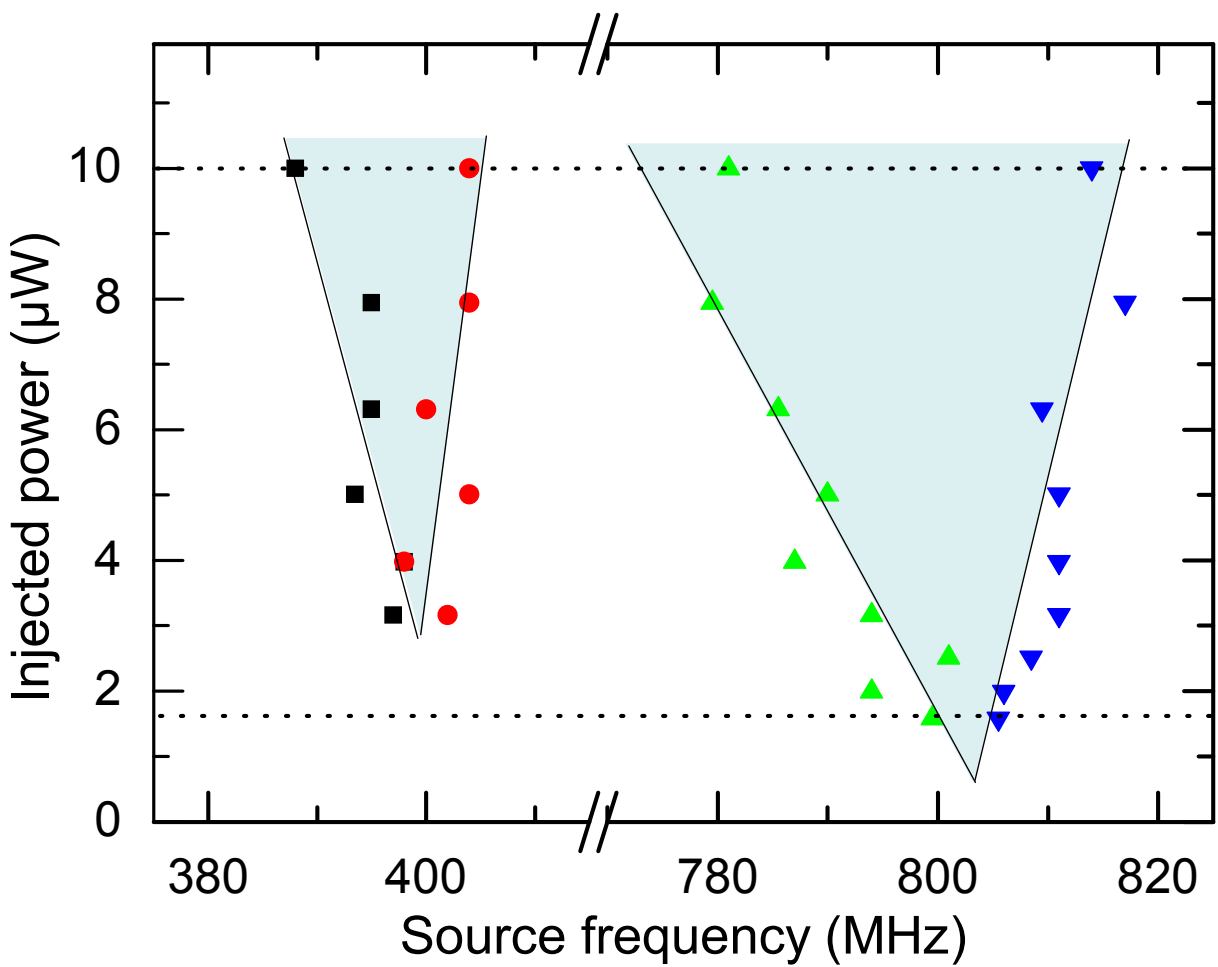} 
	\caption{(Color online) Arnold tongues : limits of the locking range of the oscillator as a function of the injected power. This graph is constructed from a series of measurements similar to those reported in Fig.~\ref{fig:locking} with varying the power of the excitation. The measurements of Fig.~\ref{fig:locking} correspond to an injected power of $10\mu W$ and $1.6\mu W$. Solid lines are guide to the eye.}
	\label{fig:Arnold}
\end{figure}

Now consider the full equation (Eq.~\ref{mathieu}) when the ac-current is injected into the oscillator. We recognize the equation of a 2D parametric oscillator, similar to a damped Mathieu equation. When $\omega_{ext}=2\omega_0$ (i.e.\ $\omega=\omega_0$), this equation predicts that a dynamical instability, defined as an exponential divergence of the orbit radius, can be reached for a small non-zero value of the excitation in the presence of damping. In our model, the confinement potential prevents the divergence and a stable periodic motion can be obtained, as experimentally observed. Other similar cases with an increasingly larger excitation threshold also exist when $\omega = \omega_0/n$ with $n$ being an integer i.e.\ when $\omega_{ext}=2 \omega=2 \omega_0/n$. This has also been observed experimentally : for $\omega_{ext}=\omega_0$ ($n=2$), a larger power is necessary to lock the peak and for $\omega_{ext}=\omega_0/2$ ($n=4$), no efficient locking could be observed. These observations are clearly seen in Fig.~\ref{fig:Arnold} which represent the Arnold tongues \cite{Pikovsky2005}. This plot is obtained in the following way: for various power of the injected microwave, the locking range is measured and the minimum and maximum frequencies of the locking range are reported in the graph. The V-shape regions thus obtained (shaded area in Fig.~\ref{fig:Arnold}) define areas where synchronization is possible. This plot clearly shows that synchronization at twice the fundamental frequency is obtained with a much smaller excitation compared to synchronization at the fundamental frequency itself. Our model also shows that shifting $\omega_{ext}$ from the $2 \omega_0 /n$ values must be compensated by a larger power: in fact, we observed that excitation at $3 \omega_0$ requires a larger power and excitations at $3/2 \omega_0$ or $4 \omega_0$ were ineffective within the experimental power range.\\ 

\begin{figure}[htb]
	\includegraphics[width=\columnwidth]{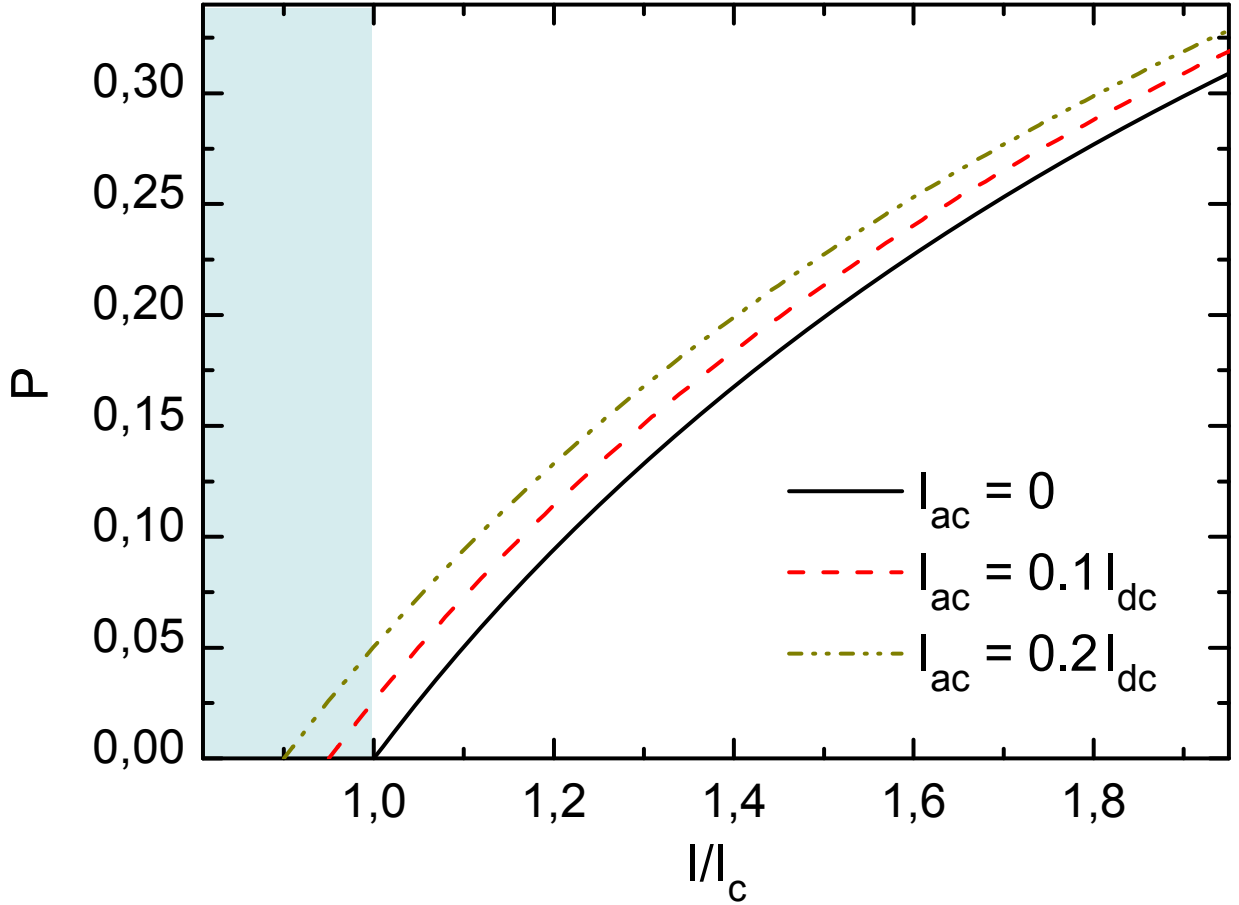} 
	\caption{(Color online) Output power as a function of the bias dc-current for different values of the ac-current $I_{ac}$. Currents are normalized to the critical current defined as the minimum current value necessary to start the vortex oscillations. For sub-critical currents (shaded area), the model predicts that an microwave excitation can switch on the oscillator.}
	\label{fig:theory}
\end{figure}

Finally it is possible to find the stability condition of Eq.~(\ref{mathieu}):
\begin{equation}
\left[M \omega^2+ G_0 \omega - K f(R)\right]^2+\left[\eta \omega - \lambda I_{dc}\right]^2 - \frac{\lambda^2 I_{ac}^2}{4}=0 
\end{equation}
This formula is obtained in the limit range where the dc-current is much larger than the ac-current i.e.\ when $I_{dc} >>I_{ac}$, assuming a circular orbit of the vortex core. From this equation, it is possible to estimate the output power of the oscillator (supposed proportional to the orbit surface $R^2$) as a function of the applied dc-current $I_{dc}$ for different values of the ac-current $I_{ac}$ (Fig.~\ref{fig:theory}). It is observed that the oscillator output power $P$ rises with dc-current due to an increase of spin transfer torque. It is also enhanced by the microwave excitation as expected for a parametric oscillator. In particular, for subcritical dc-current, the model predicts that sufficiently large ac-current could start vortex oscillations \cite{UrazhdinParam2010}.

\section{Conclusion}
In this paper we demonstrated that low RA MTJ subjected to low in-plane field and high current are powerful vortex oscillators. The vortex oscillation is ascribed to spin transfer from a non uniform polarizer onto the vortex core in the free layer. This vortex oscillator is particularly efficiently synchronized by an ac current which frequency is twice the oscillator natural frequency. All our observations on synchronization can be described in the framework of a confined 2D parametric oscillator. Parametric amplification below the critical current is also expected from the model and has to be checked experimentally.\\

\section{Acknowledgements} 
We thank Y. Liu and M. Dovek from Headway Technology (Milpitas, CA, USA) for providing us with low RA tunnel junctions. Helpful discussions with A. Thiaville, B. Canals and E. Bonet are gratefully acknowledged. This work was partially supported by the ERC Advanced Grant HYMAGINE.

\bibliographystyle{apsrev}

\end{document}